# To aggregate or not to aggregate: Forecasting of finite autocorrelated demand

Bahman Rostami-Tabar[1] [*], M. Zied Babai [2], Aris Syntetos [1]

[1] Cardiff Business School, Cardiff University, Cardiff, United Kingdom
[2] Kedge Business School, Talence, France

## Abstract

Temporal aggregation is an intuitively appealing approach to deal with demand uncertainty. There are two types of temporal aggregation: non-overlapping and overlapping. Most of the supply chain forecasting literature has focused so far on the former and there is no research that analyses the latter for auto-correlated demands. In addition, most of the analytical research to-date assumes infinite demand series' lengths whereas, in practice, forecasting is based on finite demand histories. The length of the demand history is an important determinant of the comparative performance of the two approaches but has not been given sufficient attention in the literature. In this paper we examine the effectiveness of temporal aggregation for forecasting finite auto-correlated demand. We do so by means of an analytical study of the forecast accuracy of aggregation and non-aggregation approaches based on mean squared error. We complement this with a numerical analysis to explore the impact of demand parameters and the length of the series on (comparative) performance. We also conduct an empirical evaluation to validate the analytical results using monthly time series of the M4-competition dataset. We find the degree of auto-correlation, the forecast horizon and the length of the series to be important determinants of forecast accuracy. We discuss the merits of each approach and highlight their implications for real world practices.

**Keywords:** Temporal Aggregation; Time Series Forecasting; Auto-Correlated Demand; Exponential Smoothing

[1] *Correspondence: B. Rostami-Tabar, Cardiff Business School, Cardiff University, Cardiff, CF10 3EU, UK. Tel.: +44-2920-870723 Email address: rostami-tabarb@cardiff.ac.uk

# 1. Introduction

Most organisations need a forecast of future demand over some planning horizon to support supply chain operation decisions (Schoenmeyr & Graves, 2009). Demand forecasting performance is subject to the uncertainty underlying the time series an organisation is dealing with. The existence of high variability in time series demand for both fast and slow-moving items poses considerable difficulties in terms of forecasting and stock control resulting in high organisational costs (Park et al., 2018; Sanders & Graman, 2009). There are many approaches that may be used to reduce demand uncertainty (Kouvelis et al., 2006) and thus improve forecasting performance. An intuitively appealing approach is to aggregate time series in lower-frequency 'time buckets'. The approach concerned is referred to as Temporal Aggregation (TA) (Amemiya & Wu, 1972). In temporal aggregation, a low frequency time series (e.g. quarterly) is derived from a high frequency one (e.g. monthly) (Nikolopoulos et al., 2011). This is achieved through the summation (bucketing) of every $m$ periods of the high frequency time series, where $m$ is the aggregation level.

There are two different types of temporal aggregation: non-overlapping and overlapping. In the former case, the time series are divided into consecutive non-overlapping buckets of time where the length of the time bucket equals the aggregation level. The aggregate demand is created by summing up the values inside each bucket. The number of aggregate periods is [$N/m$], where $N$ is the number of the original periods, $m$ the aggregation level and the operator [$x$] returns the integer part of $x$. Consequently, the number of time periods in the aggregate demand time series is less than that corresponding to the original demand time series. The overlapping case is somewhat different in that it resembles the mechanism of a moving window technique where the window's size equals the aggregation level. At each period, the window is moved one step ahead, so the oldest observation is dropped and the newest is included. It is observed that the number of overlapping aggregated periods is higher than those of the non-overlapping case and is equal to *N-m*+1. Therefore, the information loss is negligible as compared to the non-overlapping case. This is an important observation in terms of data availability which may have considerable implications on the forecast accuracy for the cases where little history of data is available. A disadvantage of the overlapping TA is that the first and last observations in the original series are under-represented in the aggregated series.

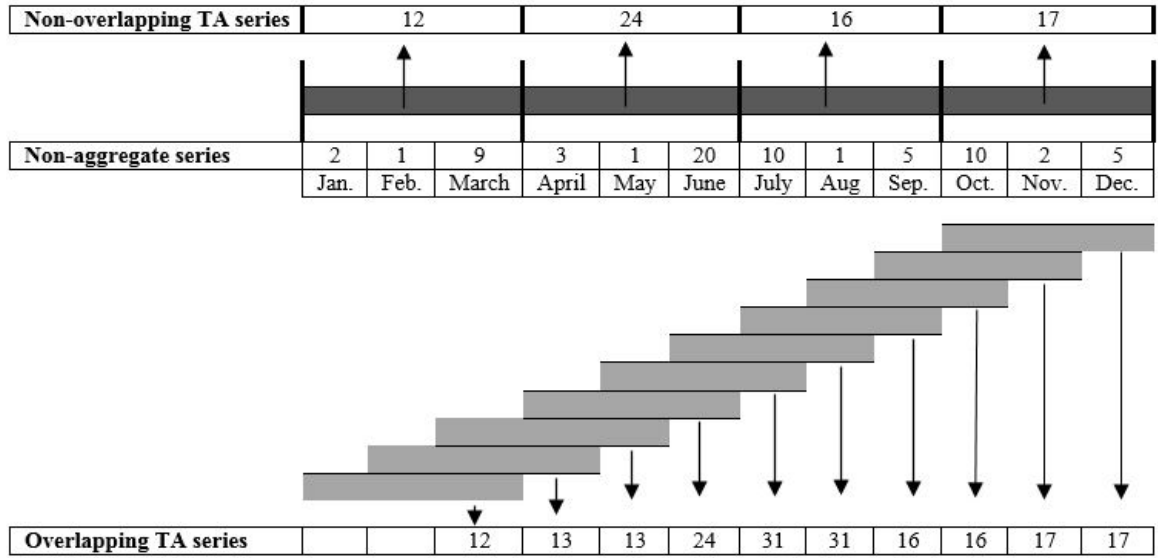

Figure 1 : An example to illustrate how a non-aggregated series is transformed into overlapping and non-overlapping temporally aggregated series.

The literature that deals with the impact of non-overlapping temporal aggregation on demand forecasting has been growing rather rapidly during the last decade (Syntetos et al., 2016). Non-overlapping aggregated demand processes have been theoretically analysed and it has been shown that the aggregation approach can improve forecast accuracy as compared to the non-aggregation approach, i.e. an approach that utilises the original series. Rostami-Tabar et al. (2013; 2014) analysed the impact of temporal aggregation on the forecast accuracy for auto-correlated stationary ARMA(1,1) demand processes where the Single Exponential Smoothing (SES) method is used to forecast demand. They showed that the forecast accuracy benefit from using non-overlapping temporal aggregation depends on the autocorrelation parameter, the aggregation level and the smoothing constant used. Furthermore, they demonstrated that for high levels of positive autocorrelation in the original series, the non-aggregation approach outperforms the aggregation one.

Unlike the fast-growing area of non-overlapping TA, the effects of overlapping TA have been rather neglected in the supply chain forecasting literature. To the best of our knowledge, there is no research that has analytically investigated the performance of the overlapping aggregation approach for auto-correlated demands. In addition, most of the analytical research assumes infinite demand series' lengths whereas, in practice, forecasting is based on finite demand information (Akcay et al., 2011). Despite the fact that the length of the series is one of the key determinants of the performance of aggregation, neither the overlapping nor the non-overlapping aggregation approach have been analysed under finite demand history lengths. We address these issues in this paper.

We do so by considering ARMA(1,1) demand processes in conjunction with the SES forecasting method. We analyse the performance of the three approaches: non-aggregation, non-overlapping aggregation, and overlapping aggregation - when the length of the demand series is finite. The use of the ARMA(1,1) demand process has motivation both from a theoretical and a practical point of view. Although many popular non-stationary times series processes are not captured by an ARMA structure (such as series with seasonality, for example), many studies have argued for its relevance in supply chain and inventory forecasting (Ali et al., 2012; Alwan et al., 2003; Chen et al., 2000; Lee et al., 2000; Rostami-Tabar et al., 2014). Hosoda et al. (2008) show real supply chain contexts where retailers and suppliers follow autoregressive order one, AR(1) and ARMA(1,1) demand processes. Disney et al. (2006) also show that the demand processes for Procter & Gamble products can be modelled as an ARMA(1, 1).

Note that our analysis is complemented by an empirical investigation where real demand data have also been considered.

We have first considered using an optimal minimum mean squared error (MMSE) forecasting method for the underlying ARMA(1,1) process; however under the MMSE method the forecast at any period depends only on the last observation (Rostami-Tabar et al., 2019) and consequently the impact of the length of series on the performance of the aggregation approaches cannot be analysed. Hence, using SES as forecasting method instead of an optimal method enables us to analyse the effect of temporal aggregation on forecast accuracy as a function of the length of series. Further, from a practical perspective, the incompatibility of SES with a stationary demand framework is less of an issue as SES is known to be used in such contexts (Babai et al., 2014). In fact, SES can adapt to changes in the level of time series by choosing appropriately the smoothing constants, which make it useful not only for stationary but also non-stationary series.

Moreover, it is worth pointing out that SES is a very popular forecasting method in industry. It is being applied in very many companies and most managers are familiar with that due to its simplicity and yet very robust performance. SES is also unbiased for the ARMA (1,1) demand process (Acar & Gardner Jr, 2012; Gardner, 1990; Kim & Ryan, 2003).

The objective of our study is to gain insights on the relative performance of the three approaches and to determine the conditions under which each approach leads to more accurate forecasts. Performance is measured by means of the Mean Squared Error (MSE). Our contribution is three-fold:

1) We derive analytical MSE expressions under the three approaches when a finite history length is used.

2) We numerically evaluate and compare the performance of the three approaches by analysing the impact of the length of the series, the aggregation level and the process autocorrelation on the forecast performance.

3) Using monthly time series from the M4 competition, we empirically evaluate and compare the performance of the three approaches.

The aggregation level is conveniently chosen to match the forecast horizon, and this makes a lot of sense from a practical perspective. Our analysis refers to the aggregated series; while generating and evaluating forecasts in the original frequency might be useful (Nikolopoulos et al., 2011), this is not covered in this paper but rather is introduced as a next step of research.

We find the degree of auto-correlation, the forecast horizon and the length of the series to be important determinants of (comparative) forecast accuracy performance. We discuss the merits of each approach and highlight their implications for real world practices. It is important to note that temporal aggregation is currently supported by very few software packages (e.g., SAS and R) and most the commercial forecasting software packages do not support it. The lack of operationalised rules that may allow switching from one approach to another goes some way to explaining why this is the case, and our research sheds light on this area.

The remainder of the paper is organised as follows. In Section 2, the relevant literature is reviewed. In Section 3, the presentation of the assumptions and the analytical expressions of the MSE, related to the three approaches, are derived. Section 4 presents the numerical results and discusses the findings of our work followed by an empirical investigation in Section 5. The conclusions, implications and suggestions for future work are presented in Section 6.

## 2. Research background

A considerable part of the supply chain forecasting literature is devoted to the non-overlapping temporal aggregation approach. The potential forecasting benefit of non-overlapping temporal aggregation in the context of supply chain was initially recognised by Willemain et al. (1994) for intermittent demands. Nikolopoulos et al. (2011) have also shown that an aggregation approach may offer considerable improvements in forecasting and stock control. They have empirically analysed the effects of non-overlapping temporal aggregation on forecasting intermittent demand requirements. Their main motivation was to reduce the number of zeros present in the original intermittent series and then forecast the series with conventional forecasting methods, once the intermittency has been reduced substantially. This work showed that the proposed methodology may indeed offer considerable improvements in terms of forecast accuracy. Since then, such findings have been further (empirically) confirmed by Babai et al. (2012) and Petropoulos and Kourentzes (2015). Spithourakis et al. (2012) extended the application of Nikolopoulos et al. (2011) to fast-moving demand data. Their results supported the forecast accuracy improvement gains obtained by temporally aggregating demand. In a fast-moving demand context under an Auto-Regressive Moving Average (ARMA) framework of analysis, Rostami-Tabar et al. (2013, 2014) considered analytically the effect of non-overlapping temporal aggregation on demand forecasting. Assuming an ARMA (1,1) demand process and Single Exponential Smoothing (SES) forecasting method, they analytically showed that, for high values of positive autocorrelation in the non-aggregated demand series, the non-overlapping temporal aggregation approach is outperformed by the non-aggregation. Moreover, they showed that the benefits of using non-overlapping temporal aggregation on the forecast accuracy depend on three factors: *i*) autoregressive and moving average parameter, *ii*) the aggregation level and *iii*) the smoothing constant. Additionally, the performance of aggregation was generally found to improve as the aggregation level increases. Instead of focusing on the optimal level of aggregation, some studies investigate the use of multiple level of temporal aggregation. In these studies, forecasts generated at multiple levels of aggregation are combined. Findings show that combining forecasts leads to more accurate forecasts as it accounts for information available at all levels and it may also help to reduce uncertainty (Kourentzes et al., 2014; Athanasopoulos et al., 2017). Kourentzes et al. (2017) contrasted the effectiveness of using a multiple aggregation level or a single optimal aggregation level (Rostami-Tabar et al., 2014) in forecast accuracy improvement. They concluded that using non-overlapping TA is beneficial to forecast performance improvement compared to the non-aggregation approach and argued that further research is needed towards identifying the optimal aggregation level. This constitutes one of the objectives of this work. It should be noted that all the above studies have analysed the performance of the non-overlapping approach under an infinite length of demand history and none of them has analysed the impact of the length on the performance which is addressed in this paper.

With regard to overlapping TA, the literature is scarcer. Mohammadipour and Boylan (2012) have analysed the theoretical properties of integer ARMA (INARMA) demand processes under the overlapping temporal aggregation approach and they showed that the aggregation of an INARMA process over a given horizon results in an INARMA process as well. Porras and Dekker (2008) have shown, based on an empirical investigation conducted with a Dutch petrochemical complex, the stock control outperformance of the overlapping temporal aggregation when compared to a bootstrapping forecasting approach proposed in Willemain et al. (1994) and to the Poisson-based stock control approach (Silver et al., 2017). More recently, Boylan and Babai (2016) have analytically compared the variance estimates of overlapping and non-overlapping aggregation for i.i.d demand processes. By means of numerical and empirical analysis, they have shown that unless the demand history is short, there is a clear advantage of using the former. However, they emphasised i.i.d demand, i.e. the empirically plausible case of auto-correlated demands has not been addressed. In this paper, we analyse the performance of overlapping temporal aggregation in the case of auto-correlated demands whilst also considering the impact of the length of the demand history on forecast performance. Moreover, its performance is examined against the non-overlapping temporal aggregation and the non-aggregation approaches.

## 3. Forecast accuracy under a finite demand history length

In this section, we examine the effectiveness of the non-aggregation (NA), non-overlapping (NOA) and overlapping (OA) TA approaches on forecast accuracy when *i*) a finite number of observations is available *ii*) SES forecasting method is used and *iii*) a forecast over a horizon *m,* is required.

We assume that the non-aggregated series $d_t$, follows an autoregressive moving average process of order (1,1) - ARMA(1,1) - that can be mathematically written by (1).

$$d_t = C + \epsilon_t + \phi d_{t-1} - \theta \epsilon_{t-1} \text{ where } |\theta| \leq 1, |\phi| \leq 1, \tag{1}$$

where $\varepsilon_t$ is the independent random variable for the non-aggregated series in period *t*, normally distributed with zero mean and variance $\sigma^2$, $\theta$ is the moving average parameter and $\phi$ is the autoregressive parameter of the non-aggregated series and *C* is a constant value. When demand series follow an ARMA(1,1) process, the auto-covariance is (Box et al., 2015):

$$\gamma_k = Cov(d_t, d_{t-k}) = \begin{cases} \dfrac{(1-2\phi\theta+\theta^2)}{1-\phi^2}\sigma^2 & k = 0 \\ \dfrac{(\phi-\theta)(1-\phi\theta)}{1-\phi^2}\sigma^2 & |k| = 1 \\ \phi^{k-1}\gamma_1 & |k| > 1 \end{cases} \tag{2}$$

We analyse the effectiveness of three approaches to produce the cumulative forecast over horizon *m* for periods *t, t+1, …, t+m-1*. The forecast is evaluated against the cumulative (aggregated) demand written as follows:

$$D_T = d_t + d_{t+1} + \cdots + d_{t+m-1} \tag{3}$$

The effectiveness of these approaches on forecast accuracy improvement is evaluated using the MSE measure. Therefore, MSE expressions of cumulative forecast over horizon *m* need to be derived for three approaches. The MSE of forecast horizon is related to the safety stock required in the inventory system, and thus has a direct cost implication.

We first derive the MSE utilising the non-aggregated demand series $d_t$. To do so, SES is applied to the non-aggregated series to produce $m$-steps-ahead forecast for periods $t, t+1, \ldots, t+m-1$. Next, we sum forecasts to obtain the cumulative forecast over horizon $m$, $f_t^m$. We assume $f_t^m = mf_t$, which is a reasonable assumption due to the stationarity of demand series and then calculate the MSE of forecasts for the non-aggregation approach as follows:

$$MSE_{NA} = var(D_T - f_t^m) = var(D_T - mf_t), \qquad (4)$$

In order to calculate the MSEs of non-overlapping and overlapping aggregation approaches, firstly buckets of aggregated series are created based on the aggregation level; then, the SES method is applied to the aggregated series to produce the cumulative forecasts over horizon $m$. The MSEs of forecast over horizon $m$ for the non-overlapping, $MSE_{NOA}$ and overlapping, $MSE_{OA}$ aggregated series are calculated as following:

$$MSE_{NOA} = var(D_T - F_{T,NOA}^1), \qquad (5)$$

$$MSE_{OA} = var(D_T - F_{T,OA}^1). \qquad (6)$$

where $F_{T,NOA}^1$ and $F_{T,OA}^1$ are the forecasts of non-overlapping and overlapping aggregated demand series in period T respectively; the forecast produced in $T-1$ for the demand in $T$, which is represented by (3). In sub-sections 3.1, 3.2 and 3.3 we derive the MSE forecast expressions of SES utilising non-aggregated, non-overlapping and overlapping temporally aggregated series.

### 3.1. Forecast accuracy using non-aggregated series

Following the Equation (4), the theoretical MSE of the non-aggregated series, can be written as follows:

$$MSE_{NA} = var(D_T - mf_t) = Var(D_T) + m^2 Var(f_t) - 2mCov(D_T, f_t), \qquad (7)$$

Subsequently, the three parts in each of the expression (7), need to be determined: *i*) variance of aggregated demand series, *ii*) variance of the forecast utilising non-aggregated series, and *iii*) the covariance between the aggregated series and the forecast utilising non-aggregated series.

Rostami-Tabar et al. (2014) showed that the variance of the non-overlapping aggregated series for an ARMA(1,1) process, $\gamma_0' = \text{var}(D_T)$ is calculated as follows:

$$\gamma_0' = m\gamma_0 + \gamma_1 \left( \sum_{k=1}^{m-1} 2(m-k)\phi^{k-1} \right) \tag{8}$$

Using SES, the forecast of series in period *t* produced at the end of period *t-1* is:

$$f_t = \sum_{k=1}^{N} \alpha(1-\alpha)^{k-1} d_{t-k} + (1-\alpha)^N f_0, \tag{9}$$

where $d_{t-k}$ is the non-aggregated series, $f_0$, is the initial forecast and $0 < \alpha \leq 1$ is the smoothing constant used in SES for the non-aggregated series. Initial forecasts play an important role in generating forecasts in all periods. They can have an important effect when the time series are short, and the smoothing constant weight is small. A common approach is to set the initial value to the first observation $f_0 = d_{t-N}$.

By considering $var(D_T) = \gamma_0'$ and substituting (B1) and (B3) in Appendix B into (7), the MSE related to forecasts utilising the non-aggregated series is calculated as follows:

$$\begin{aligned} MSE_{NA} = {}& m\gamma_0 + \gamma_1 \left( \sum_{i=1}^{m-1} 2(m-k)\phi^{k-1} \right) + \\ & (m^2) \left( \begin{array}{l} \dfrac{\alpha\gamma_0 \left(1-(1-\alpha)^{2N}\right)}{(2-\alpha)} + 2\alpha(1-\alpha)^{2N-1}\gamma_0 + (1-\alpha)^{2N}\gamma_0 + \\ \dfrac{2\alpha\gamma_1}{(2-\alpha)} \sum_{i=1}^{N-1}(1-\alpha)^i \phi^{i-1}\left(1-(1-\alpha)^{2(N-i)}\right) + \sum_{i=1}^{N-1} 2\alpha\phi^{N-i-1}(1-\alpha)^{N+i-1}\gamma_1 \end{array} \right) \\ & -\dfrac{2\alpha\gamma_1(1-\phi^m)}{(1-\phi)(1-\phi+\alpha\phi)} \times \left(1-(\phi-\alpha\phi)^N\right) + (1-\alpha)^N \phi^{N-1}\gamma_1 \times \dfrac{1-\phi^m}{1-\phi} \end{aligned} \tag{10}$$

## 3.2. Forecast accuracy using non-overlapping aggregated series

The *MSE* related to non-overlapping aggregated series presented in Equation (5) can be expressed as follows:

$$MSE_{NOA} = var\left(D_T - F^1_{T,NOA}\right) = var(D_T) + var\left(F^1_{T,NOA}\right) - 2cov\left(D_T, F^1_{T,NOA}\right), \quad (11)$$

The variance of aggregated demand has already been calculated in Equation (8). Additionally, we need to determine *i)* the variance of the aggregated forecasts utilising the non-overlapping TA series, and *ii)* the covariance between the aggregated series and its forecast.

Using SES, the forecast of non-overlapping aggregated series in period *T* produced at the end of period *T-1* is:

$$F_{T,NOA} = \sum_{k=1}^{[N/m]} \beta_N (1-\beta_N)^{k-1} D_{T-k,NOA} + (1-\beta_N)^{[N/m]} F_{0,NOA}, \quad (12)$$

where $F_{0,NOA}$ is the initial non-overlapping TA forecast and $0 < \beta_N \leq 1$ is the smoothing constant used in SES for the non-overlapping aggregated series. The non-overlapping aggregated demand series over *m* periods, $D_{T-k,NA}$, can be expressed as a function of the non-aggregate demand series as follows:

$$D_{T-k,NOA} = \sum_{l=1}^{m} d_{t-(k-1)m-l}, \quad k = 1, 2, \cdots, [N/m] \quad (13)$$

From Rostami-Tabar et al. (2014), we have the relations between the process parameters of the aggregated and the non-aggregated series for an ARMA(1,1) process as follows:

$$\gamma'_0 = m\gamma_0 + \gamma_1 \left(\sum_{k=1}^{m-1} 2(m-k)\phi^{k-1}\right) \quad (14)$$

$$\gamma'_1 = \gamma_1 \left(\sum_{k=1}^{m} k\phi^{k-1} + \sum_{k=2}^{m} (k-1)\phi^{2m-k}\right) \quad (15)$$

$$\phi' = \phi^m \quad (16)$$

By substituting (14), (C3) and (C5) in the Appendix C into (11), the MSE of forecast utilising the non-overlapping temporally aggregated series is:

$$MSE_{NOA} = \left( m\gamma_0 + \gamma_1 \left( \sum_{k=1}^{m-1} 2(m-k)\phi^{k-1} \right) \right) \left( \frac{\beta_N \left(1-(1-\beta_N)^{2[N/m]}\right)}{(2-\beta_N)} + (1-\beta_N)^{2[N/m]} + 2\beta_N(1-\beta_N)^{2[N/m]-1} \right) +$$

$$m\gamma_0 + \gamma_1 \left( \sum_{k=1}^{m-1} 2(m-k)\phi^{k-1} \right) + \sum_{i=1}^{([N/m]-1)} \frac{2\beta_N \gamma_1}{(2-\beta_N)} \left( \frac{1-\phi^m}{1-\phi} \right)^2 (1-\beta_N)^i \phi^{m(i-1)} \left( 1-(1-\beta_N)^{2[N/m]-i} \right) \quad (17)$$

$$-2\left( \beta_N \gamma_1 \times \left( \frac{1-\phi^m}{1-\phi} \right)^2 \frac{1-(\phi^m - \beta_N \phi^m)^{[N/m]}}{(1-\phi^m + \beta_N \phi^m)} + \phi^{m([N/m]-1)}(1-\beta_N)^{2[N/m]} \left( \frac{1-\phi^m}{1-\phi} \right)^2 \gamma_1 \right).$$

### 3.3. Forecast accuracy using overlapping aggregated series

The *MSE* related to overlapping aggregated demand series presented in Equation (6) can be represented as follows:

$$MSE_{NOA} = var\left(D_T - F^1_{T,OA}\right) = var(D_T) + var\left(F^1_{T,OA}\right) - 2cov\left(D_T, F^1_{T,OA}\right), \quad (18)$$

In addition to the variance of aggregated series calculated in Equation (8), we need to determine *i*) the variance of the aggregated forecast utilising the overlapping TA series, and *ii*) the covariance between the aggregated demand series and its forecast.

Using SES, the forecast of overlapping aggregated demand in period *T* produced at the end of period *T-1* is:

$$F^1_{T,OA} = \sum_{k=1}^{N-m+1} \beta_o (1-\beta_o)^{k-1} D_{T-k,OA} + (1-\beta_o)^{N-m+1} F_{0,OA}, \quad (19)$$

where $F_{0,OA} = D_{T-(N-m+1)}$, is initial forecast and $0 < \beta_O \leq 1$ is the smoothing constant used in SES method for the overlapping aggregated series.

We need to determine the autocovariance of the overlapping aggregated process and its relationship with the non-aggregate process to calculate the variance of the overlapping aggregated forecast and its covariance with the overlapping aggregated demand series. Having *N* demand observations at the end of period *t-1* in the non-aggregated series, the overlapping aggregated demand series over *m* periods, $D_{T-k,OA}$, can be expressed as follows:

$$D_{T-k,OA} = \sum_{l=1}^{m} d_{t-(l+k-1)} \quad k = 1,2,...,N-m+1 \tag{20}$$

The MSE derivation of the overlapping aggregation approach requires the calculation of the auto-covariance function of the aggregated process, which, to the best of our knowledge, has never been performed before in the literature. The auto-covariance of lag $k \geq 0$, for the overlapping aggregated series is calculated as follows:

$$\gamma_k'' = cov\left(\sum_{l_1=1}^{m} d_{t-(l_1-1)}, \sum_{l_2=1}^{m} d_{t-(l_2+k-1)}\right) = \sum_{l_1=1}^{m}\sum_{l_2=1}^{m} cov\left(d_{t-(l_1-1)}, d_{t-(l_2+k-1)}\right) \tag{21}$$

By substituting (D7) in Appendix D, (E5) in Appendix E and (F3) in Appendix F into (18), the MSE of forecast utilising the overlapping aggregated series is:

$$MSE_{OA} = m\gamma_0 + 2\gamma_1 \sum_{i=1}^{m-1}(m-i)\phi^{i-1} +$$

$$\frac{\beta_o\left(m\gamma_0 + 2\gamma_1 \sum_{i=1}^{m-1}(m-i)\phi\right)\left(1-\left((1-\beta_o)^2\right)^{N-m+1}\right)}{(2-\beta_o)} - 2\beta_O\gamma_1 \times \left(\frac{1-\phi^m}{1-\phi}\right)^2 \left(\frac{1-(\phi(1-\beta_O))^{N-m+1}}{1-\phi+\phi\beta_O}\right) \tag{22}$$

$$+\frac{2\beta_o}{(2-\beta_o)}\sum_{k=1}^{N-m}(1-\beta_o)^k \left(\sum_{\substack{i=1 \\ \forall |k+j-i|=0}}^{m}\sum_{j=1}^{m}\gamma_0 + \sum_{\substack{i=1 \\ \forall |k+j-i|=1}}^{m}\sum_{j=1}^{m}\gamma_1 + \sum_{\substack{i=1 \\ \forall |k+j-i|>1}}^{m}\sum_{j=1}^{m}\phi^{(|k+j-i|-1)}\gamma_1\right)\left(1-\left((1-\beta_o)\right)^{2(N-m+1-k)}\right),$$

Due to the complexity of the MSE expressions given by (10), (17) and (22), deriving mathematical proofs to determine the conditions under which each approach provides more accurate forecast is not feasible. Therefore, a numerical analysis will be conducted in Section 4 to examine the impact of the process parameters, the aggregation level and the history length on the comparative performance of the three approaches. The MSEs derived in section 3 are available (upon request also from the corresponding author) in R software to enable the reproducibility of our results (Boylan, 2016).

## 4. Numerical analysis and discussion

In this section, the effectiveness of the non-aggregation, overlapping and non-overlapping temporal aggregation approaches are assessed by analysing the ratio of their *MSE*s. The ratio values of $MSE_{NOA} / MSE_{NA}$ and $MSE_{OA} / MSE_{NA}$ show the pair-wise comparative performance of these approaches. If the ratio is greater than one, it means that the aggregation approach does not improve forecast accuracy. In addition to the MSE comparisons, the effect of autocorrelation through the autoregressive and moving average process parameters (i.e. $\phi$ and $\theta$), aggregation level ($m$) and the length of series ($N$) on the MSE is analysed and the superiority conditions of each approach are determined.

Given that the autocorrelation of the demand process is one of the key factors impacting the performance of the non-overlapping TA approach (Rostami-Tabar et al., 2013; 2014), in this section, we discuss the results of the numerical investigation in four cases, corresponding to conditions where the autocorrelation of the ARMA(1,1) demand process is: *i)* negative, *ii)* positive *iii)* oscillating between positive and negative values depending on its lag and iv) zero which corresponds to a white noise process.

**Figure** 2 illustrates the four categories discussed above for the ARMA(1,1) process.

Given the considerable number of control and process parameter combinations, it is natural that only some results may be presented here. Our analysis has been performed for all variations of parameters, but only the below discussed cases are presented in the paper. The numerical analysis output was judged to be represented sufficiently through the consideration of $m = 2$, 7, 12, $\phi = -0.95: +0.95$ (with an increment step of 0.05), $\theta = -0.95: +0.95$ (with an increment step of 0.05) and $N = 24, 36, 48, 60, 72, 84, 108, 132, 154, 250, 500$. Small and large $N$ values can represent short and long-time series, respectively.

We report results by considering the smoothing constant parameters for SES that provides the minimum MSE in equations (10), (17) and (22). We also provide in the Appendix G results for a smoothing constant equal to 0.1, 0.2 and 0.3, to see the deviation of the results when a non-smoothing constant is used. It should be noted that although the results are reported for these particular values of the smoothing constant, one may choose any other values between zero and one, as the MSE expressions are functions of the smoothing constant.

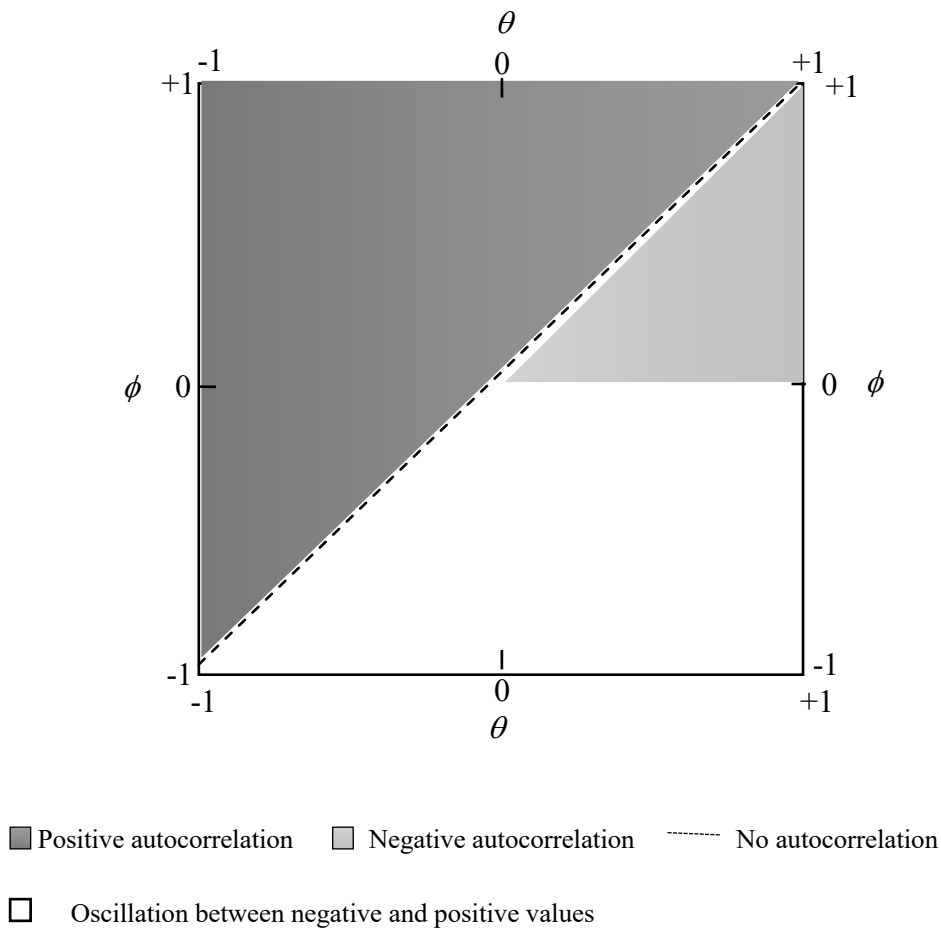

Figure 2 : Autocorrelation associated with an ARMA(1,1) process

As shown in

Figure 2, there are various auto-correlation cases corresponding to combinations of $\phi$ and $\theta$. We calculate the MSEs corresponding to each combination of $\phi$ and $\theta$ for each case and for the given values of *N* and *m*. We then determine the arithmetic average of MSEs across all process parameter combinations. Next, we calculate the ratio values of $MSE_{NOA} / MSE_{NA}$ and $MSE_{OA} / MSE_{NA}$ for each case. Finally, we repeat it for all *N* and *m* values and present results in Figure 3. In each graph, the *y*-axis represents MSE ratios and the *x*-axis indicates the number of observations. The figure shows the impact of the aggregation level, *m* and number of observations, *N* on the ratios $MSE_{NOA}/MSE_{NA}$ and $MSE_{OA}/MSE_{NA}$ for each autocorrelation case shown in

Figure 2.

Figure 3a shows the MSE ratio results corresponding to negative autocorrelation. Negative auto-correlation implies that small (large) values tend to be followed by large (small) values which causes fluctuations and variability in the time series. We are interested in determining the conditions under which the ratio is less than one, meaning that TA improves forecast accuracy. The results indicate that the ratios are always less than one, i.e. temporally aggregating demand always leads to better results than using the original series. The rate of improvement may reach 34% for overlapping TA and 32% for the non-overlapping TA case. Moreover, the results show that the overlapping TA clearly outperforms the non-overlapping one. However, by increasing the length of series, the difference between their performance becomes negligible.

The results also show that for a longer forecast horizon (i.e. for higher value of aggregation level *m*), the comparative forecast accuracy performance of both overlapping and non-overlapping TA increases. The former outperforms the latter except for high number of observations in which they perform similarly. The outperformance of the overlapping TA (to both the non-overlapping and non-aggregation approaches) becomes considerable when the length of the series is short. As shown in Figures G1-G3 in the Appendix, the outperformance of the overlapping approach for short series compared to the non-overlapping one is pronounced when a non-optimal smoothing constant is used. This is expected since for short histories, the loss of information under the non-overlapping TA should be associated with a loss in forecast accuracy too. The results presented in Figure 3b correspond to the conditions where the autocorrelation of different lags oscillates between positive and negative values for the non-aggregated series. Results are similar to the case of negative autocorrelation depicted in the Figure 3a. Figure 3b shows that TA improves forecast accuracy and it performs better for longer forecast horizons, and overlapping TA provides forecasts that are more accurate than non-aggregation and non-overlapping TA. As the length of the series increases, the performance of NOA and OA approaches becomes similar and for very lengthy series they eventually perform the same.

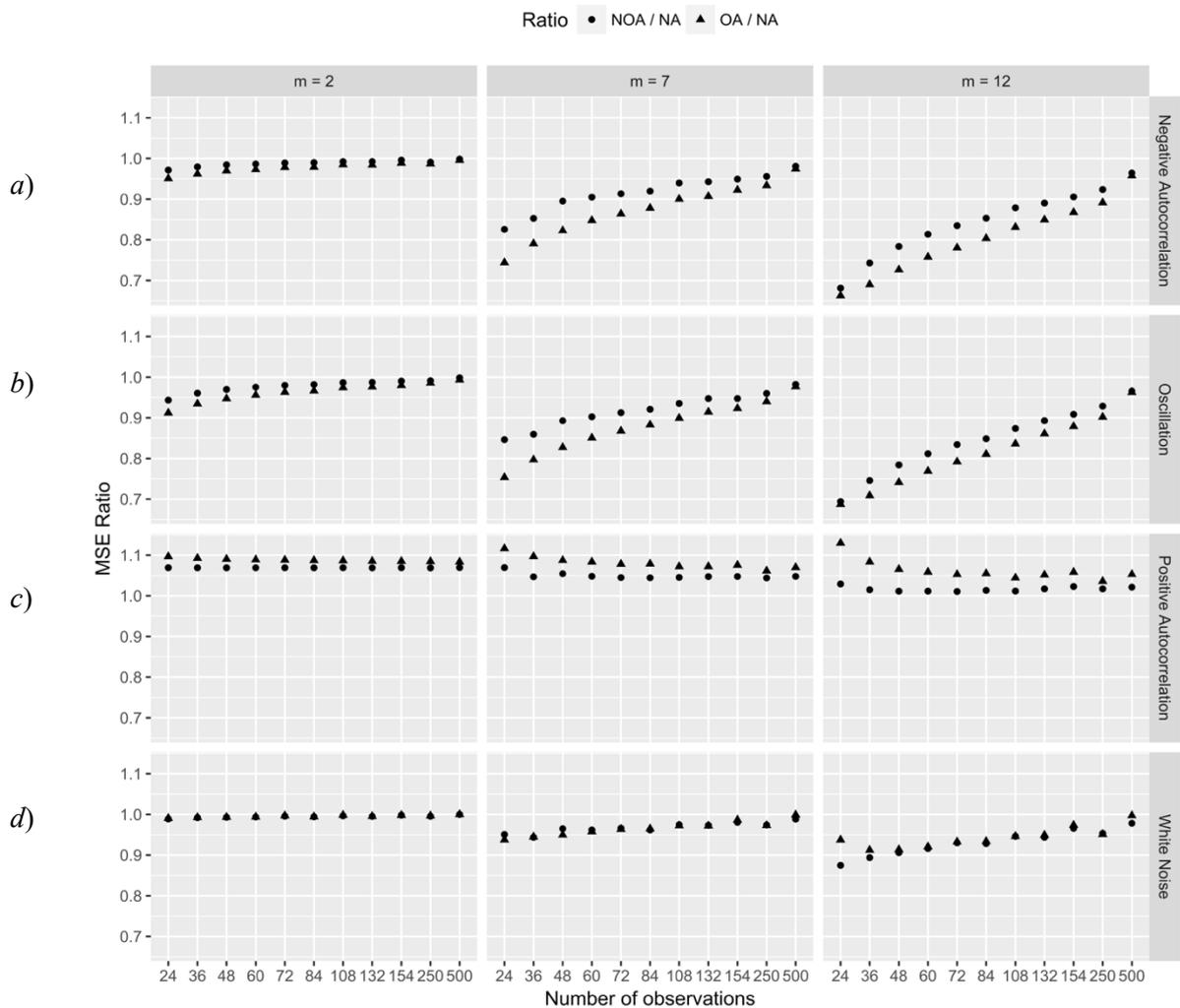

Figure 3: Ratio of MSEs for non-overlapping and overlapping aggregation to the non-aggregation approach for an ARMA(1,1) process with (a) negative autocorrelation, (b) autocorrelation oscillating between positive and negative, (c) positive autocorrelation and (d) no autocorrelation/white noise.

It should be noted that when the autocorrelation is negative or fluctuates between positive and negative vales (the non-aggregated series oscillate around a constant value), SES with a low smoothing constant value tends to behave like an average method. Obviously, by initializing SES with one observation, SES may react slowly to random fluctuations to reach the overall average of the series. However, when TA is used, the initial value represents a sum of $m$ non-aggregated demands. This reduces the random fluctuations in the aggregated series and, therefore, makes SES reaching faster the overall average compared to the case with the non-aggregated series, which ultimately improves the forecast accuracy. On the other hand, when the autocorrelation is positive, the series often behave like a random walk and SES tends to work more with a high value of the smoothing constant (it tends to work like the naïve method), which makes the temporal aggregation leading to poor performance.

The good performance of TA for smaller sample size of $N$ could be attributed to the reduction of random variations (fluctuations) in the series. Note that for series with negative autocorrelation or oscillated autocorrelation, series show random fluctuations, in comparison to series with positive autocorrelation. Therefore, using TA may reduce them and consequently improve MSE. Moreover, for lengthier series (large sample size of of $N$) the MSE converges to a constant value and including more observations does not change the MSE, therefore MSE for all approaches become very close.

Figure 3c portrays the performance of the three approaches when the autocorrelation of the demand process is always positive. Positive autocorrelation implies that small (large) demand values tend to be followed by small (large) values too, thus the relative smoothness of the resulting series.

We know from Rostami-Tabar et al. (2013, 2014) that when the autocorrelation is highly positive, non-overlapping temporal aggregation does not improve the forecast accuracy. The results in Figure 3c show that overlapping TA does not improve forecast accuracy either since the ratio $MSE_{OA}/MSE_{NA}$ is always higher than one. Such results are intuitively appealing. With high positively auto-correlated series, adjacent observations are very close in size and they follow each other, meaning that the series are smooth, and the latest observations are crucial for forecasting purposes; consequently, there is no need to use long historical observations and apply TA approaches to produce forecasts.

Finally, Figure 3d shows the MSE ratios in the absence of any autocorrelation in the series. This relates to an ARMA(1,1) process where $\phi = \theta$ which corresponds to a non-correlated (white noise) or i.i.d process. We find that the TA approach always outperforms the non-aggregation one. Ratios are close to one for lower forecast horizons ($m = 2$), but the gain of using temporally aggregated series in terms of forecast accuracy become more noticeable for higher forecast horizons and may rich up to 13%.

### 4.1. Theory-informed operationalised rules

Before we close this section, we wish to turn the results of our study into operationalised rules that could be of benefit to forecasting practice.

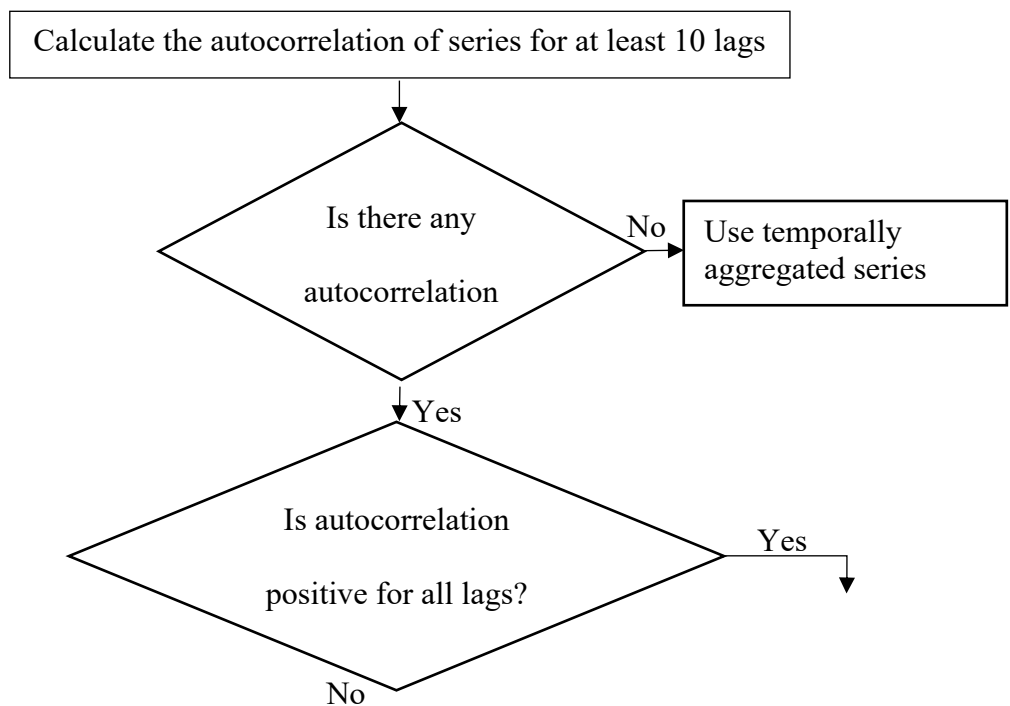

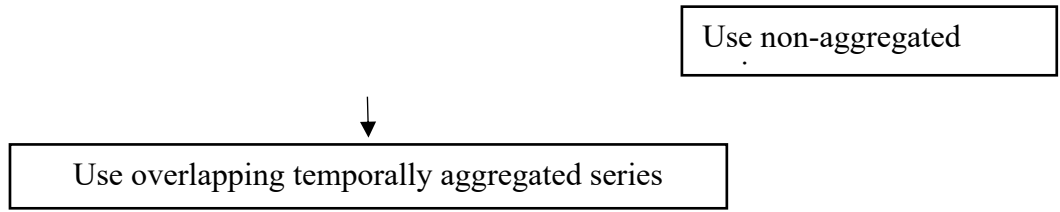

Figure 4: Rules to choose between using the original or temporally aggregated series for forecasting purposes

As already discussed in this paper, forecasters have two alternatives to produce forecasts: i) use the original series ii) use temporally aggregated series. However, and to the best of our knowledge, temporal aggregation is not included in the functionality of most of the commercial forecasting software packages. This may be attributed to the fact that the benefits of TA are not yet very clear (neither in the academic nor in the practitioner community). Even more importantly, knowledge of when to switch from one approach to the other to enable forecast accuracy improvements is lacking.

Having already analysed and discussed the potential benefits of TA, this study is the first attempt towards the derivation of theory-informed operationalised rules to enable choosing between the original or temporally aggregated series for forecasting purposes.

Figure 4 summarises the rules that can be used for operational forecasting decision making. In fact, if there is no autocorrelation, aggregated series should be used to forecast demand (the two aggregation approaches perform equally well) and in the case where the autocorrelation is not positive for all lags, overlapping temporal aggregation is the preferred approach. In all other cases, disaggregated series should be used to forecast demand. We have created a function in *R* to conduct our analysis. This may become available upon request from the corresponding author of the paper, and it should enable the reproducibility of our results (Boylan, 2016) and facilitate sensitivity analysis and extensions to the work described here.

## 5. Empirical study

In order to empirically validate the findings presented in the theoretical part, the monthly time series of the M4-Competition are used, which include 48,000 series from different domains (e.g., industry, finance, etc.). The description, and the number of time series in each category are given by Makridakis et al. (2020). Given the various lengths of the time series, the monthly M4 dataset provides us with an opportunity to analyse the impact of the time series length on the forecast accuracy. Moreover, the data is publicly available, which is an important enabler of reproducibility work (Boylan, 2016).

For the purpose of the empirical investigation, the maximum series' length is fixed to 500 periods as in the numerical part. Moreover, the series with a length less than 320 periods are excluded. The minimum length of 320 periods is considered to enable the use of at least 25 periods under non-overlapping aggregation when the aggregation level is equal to 12. This screening process results in 5,092 series, which are used for the empirical evaluation.

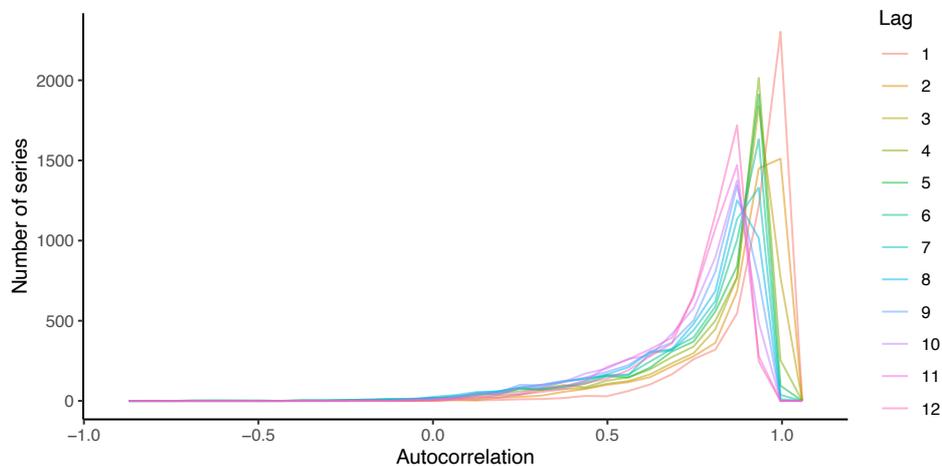

Figure 5: Distribution of the autocorrelation of lag=1 to lag= 12 for the selected monthly M4 competition series

Figure 5 illustrates the distribution of the autocorrelation values for the selected series from lag = 1 to lag = 12. The distribution of the autocorrelation lags indicates that almost all series are highly positive autocorrelated. This means that our empirical investigation considers only such cases (of highly positive autocorrelated series). This is a limitation as it enables the validation of only a part of the theoretical results discussed in Section 4. To overcome this limitation, further investigation is required on empirical datasets that cover series with negative autocorrelations and/or series where autocorrelations oscillates between positive and negative values.

We perform a rolling origin forecast evaluation using SES for each time series, each selected length, and each approach (i.e. non-overlapping aggregation, overlapping aggregation, no aggregation). To evaluate the effect of length, we consider a slide moving window, where the size of the window equals the length under consideration. To create many replications, we move the window one period ahead in each step, which includes a new observation and excludes the oldest one. Then, we create forecasts using the three approaches and calculate forecast errors. Similar to the analytical section, forecast accuracy is reported using MSE. For each approach, the average MSE is calculated through rolling origins and across all-time series.

Finally, comparative performance is evaluated by means of reporting the ratio of the MSE of (non-overlapping and overlapping) aggregation to that of non-aggregation approach. Please also note that no assumption about the underlying demand process is required here, such as the ARMA (1,1) process assumption made in the analytical part of this work.

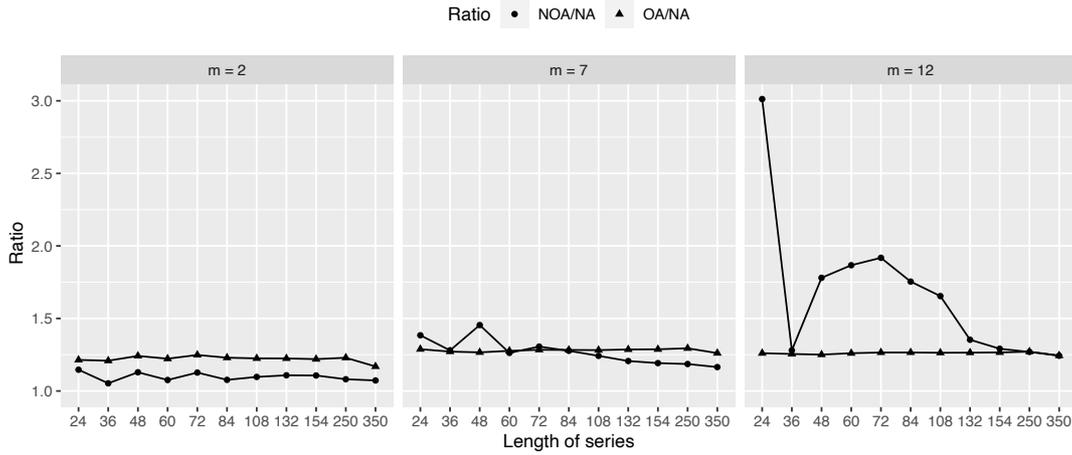

Figure 6: Ratio of empirical MSEs for non-overlapping and overlapping aggregation to the non-aggregation approach

Figure 6 illustrates the result of the empirical evaluation. Note that we only show the ratio results for series' lengths up to 350 periods to allow for a better readability of the graphs, knowing that for higher lengths the behaviour of the curves remain the same. It is evident from the results that regardless of the length of series and the level of aggregation, the MSE ratio is always greater than one. This means that the no-aggregation approach outperforms non-overlapping and overlapping aggregation. Given the characteristics of the time series (associated with high positive autocorrelation), the empirical results confirm the analytical results. The empirical results demonstrate the robustness of our theoretical findings for positively autocorrelated series, beyond the case of ARMA(1,1) demand. Note that the ratio of NOA/NA is associated with some high values for *m*=12 and N<=108. This might be due to the fact that the aggregated series using NOA result in only a few observations available to produce forecasts. In terms of the impact of the length of series and the difference between the non-aggregation and aggregation approaches, it is hard to come up with some consistent differences.

It is worth noting the importance of testing the validity of our results on time series with negative autocorrelation and/or cases where the autocorrelation oscillates between positive and negative values. While many series in the business context might be positively autocorrelated, this is not always so.

## 6. Implications, conclusion and future work

Temporal aggregation has been previously researched in the supply chain forecasting literature as a viable option to improve forecast performance. Although this would typically cover the non-overlapping case, the consideration of overlapping temporal aggregation (TA) has been neglected, for autocorrelated demand processes and a finite number of available observations, both of which are both important features of real-world demand data. However, it should be noted that consideration of non-overlapping TA, especially for higher aggregation levels, is subject to data availability. Although this might be less of an issue in modern business settings, clearly non-overlapping TA may not constitute a viable option when short demand histories are available. Tremendous recent developments in terms of computing storage capacity facilitate the accumulation of very lengthy series although we have come across situations/companies where only a few years' data is stored. In the overlapping temporal aggregation case, loss of information is less severe compared to the non-overlapping TA, which makes it an appropriate alternative even for short time series.

In this paper, we have analytically evaluated the effectiveness of the overlapping temporal aggregation approach against the non-aggregation and the non-overlapping temporal aggregation ones on forecasting performance. The objective is to generate a cumulative forecast over a horizon $m$. We assume that the non-aggregate series follow an AutoRegressive Moving Average process of order (1,1), ARMA(1,1). Moreover, we assume that the length of the demand histories is finite. Forecasting is assumed to be relying upon a Single Exponential Smoothing (SES) procedure. We have derived the MSE expressions under the three approaches to identify the conditions under which each approach outperforms the others. Given the assumptions discussed above, the main findings of this paper can be summarised as follows:

- First of all, when forecasting negatively auto-correlated series the use of the TA approach is recommended. The rate of MSE improvement may reach 34% and 32% for the overlapping and non-overlapping case, respectively.
- Second, it is recommended that for high positively autocorrelated series, the non-aggregated series should be used for forecasting purposes.
- Third, when the autocorrelation of different lags alternates between successive negative and positive autocorrelation values, TA performs well. The rate of improvement may reach 31% and 30% for the overlapping and non-overlapping case, respectively.
- Fourth, when the demand series resembles to an i.i.d process, TA outperforms the non-aggregation approach.
- Fifth, the performance of both non-overlapping and overlapping TA improves considerably when forecasting over longer horizons (e.g. $m > 6$). Therefore, as we look further in the future, it is recommended to use TA approaches.
- Finally, it is clear from the analytical results that the length of series plays an important role on the comparative effectiveness of the three approaches. When using short time series, the overlapping TA approach performs better. As more historical data is used in the forecasting process, the performance of overlapping and non-overlapping approach becomes similar. It is also observed that for all three approaches, the MSEs reduce very slowly when the length of the time series is greater than 250; the rate of MSE reduction is less than 0.02%. This may imply that there is a cutoff point in terms of historical data needed, beyond which there are no further improvements in forecast accuracy. Further investigation is needed to find the optimal cut-off point.

Given the current under-consideration of temporal aggregation in inventory forecasting software solutions and given its value as a promising uncertainty reduction time series transformation approach revealed in this study, research into any of the following areas would appear to be merited.

- Expansion of the analytical work discussed in this paper on higher order stationary processes and more importantly on non-stationary processes, with patterns including trend and seasonality (Boylan et al. 2014) is a very important issue both from an academic and practitioner perspective. Similarly, the consideration of other popular forecasting methods is an important issue too.
- The evaluation of the forecasting performance of overlapping and non-overlapping temporal aggregation at the original frequency of series should also be investigated. This may require a choice of an optimal disaggregation approach.
- Determining the optimal cut-off points of the length of series and forecast horizon for a given time series to decide when to switch from non-aggregation to temporal aggregation is an important avenue for future investigation.
- Finally, another interesting avenue for further research is to analyse the combination of the overlapping and the non-overlapping temporal aggregation on forecast accuracy performance.

## Appendix A: Autocovariance of the non-overlapping aggregated series

The covariance of non-overlapping aggregated demand and non-aggregated demand of lag *k* is calculated as follows:

$$cov(D_{T,NA}, d_{t-k}) = cov(d_t + d_{t+1} + \cdots + d_{t+m-1}, d_{t-k}) = \gamma_k + \gamma_{k+1} + \cdots + \gamma_{k+m-1} =$$
$$\phi^{k-1}\gamma_1 + \phi^k\gamma_1 + \cdots + \phi^{k+m-2}\gamma_1 = \phi^{k-1}\gamma_1\left(\frac{1-\phi^m}{1-\phi}\right), \quad k \geq 1 \qquad (A1)$$

The auto covariance of aggregated series at lag *k* is calculated as follows:

$$cov(D_{T,NOA}, D_{T-k,NA}) = cov(D_{T,NOA}, d_{t-(k-1)m-1} + \cdots + d_{t-km}) =$$
$$cov(D_{T,NOA}, d_{t-((k-1)m+1)}) + cov(D_{T,NOA}, d_{t-((k-1)m+2)}) + \cdots + cov(D_{T,NOA}, d_{t-km}) \qquad (A2)$$

By substituting (A1) into (A2), we get:

$$cov(D_{T,NA}, D_{T-k,NA}) = cov(D_{T,NOA}, d_{t-((k-1)m+1)}) + cov(D_{T,NOA}, d_{t-((k-1)m+2)}) + \cdots + cov(D_{T,NOA}, d_{t-km})$$

$$= \phi^{m(k-1)}\gamma_1 \times \left(\frac{1-\phi^m}{1-\phi}\right) + \phi^{m(k-1)+1}\gamma_1 \times \left(\frac{1-\phi^m}{1-\phi}\right) + \cdots + \phi^{mk-1}\gamma_1 \times \left(\frac{1-\phi^m}{1-\phi}\right) \quad \text{(A3)}$$

$$= \phi^{m(k-1)}\gamma_1 \times \left(\frac{1-\phi^m}{1-\phi}\right)^2$$

Therefore, the autocovariance of non-overlapping temporal aggregation series is:

$$\gamma'_k = \phi^{m(k-1)}\gamma_1 \times \left(\frac{1-\phi^m}{1-\phi}\right)^2 \quad \text{for } k \geq 1. \tag{A4}$$

## Appendix B: Variance of non-aggregated demand and the covariance of aggregated demand with non-aggregated forecast

By considering Equation (9) and the fact that $Var(d_{t-k}) = \gamma_0$ and $Cov(d_t, d_{t-k}) = \gamma_k$ for all $k \geq 1$, and then substituting (2) into it, the variance of non-aggregated forecast at period $t$, var $(f_t)$ is calculated as follows:

$$var(f_t) = var\left(\sum_{k=1}^{N}\alpha(1-\alpha)^{k-1}d_{t-k} + (1-\alpha)^N f_0\right) = \frac{\alpha\gamma_0\left(1-(1-\alpha)^{2N}\right)}{(2-\alpha)} + 2\alpha(1-\alpha)^{2N-1}\gamma_0 + (1-\alpha)^{2N}\gamma_0 +$$

$$\sum_{i=1}^{N-1}\frac{2\alpha}{(2-\alpha)}(1-\alpha)^i \gamma_i \left(1-(1-\alpha)^{2(N-i)}\right) + \sum_{i=1}^{N-1}2\alpha(1-\alpha)^{N+i-1}\gamma_{N-i}$$

$$= \frac{\alpha\gamma_0\left(1-(1-\alpha)^{2N}\right)}{(2-\alpha)} + 2\alpha(1-\alpha)^{2N-1}\gamma_0 + (1-\alpha)^{2N}\gamma_0 + \tag{B1}$$

$$\frac{2\alpha\gamma_1}{(2-\alpha)}\sum_{i=1}^{N-1}(1-\alpha)^i \phi^{i-1}\left(1-(1-\alpha)^{2(N-i)}\right) + 2\alpha\gamma_1\sum_{i=1}^{N-1}\phi^{N-i-1}(1-\alpha)^{N+i-1}$$

The covariance between the aggregated demand $D_T$ and non-aggregated forecast $f_t$ is calculated as follows:

$$cov(D_T, f_t) = cov\left(D_T, \sum_{k=1}^{N}\alpha(1-\alpha)^{k-1}d_{t-k} + (1-\alpha)^N f_0\right)$$

$$= \alpha\left(cov(D_T, d_{t-1}) + cov(D_T, (1-\alpha)d_{t-2}) + \cdots + cov(D_T, (1-\alpha)^{N-1}d_{t-N})\right) + cov(D_T, (1-\alpha)^N d_{t-N}) \tag{B2}$$

We derive the autocovariance of the non-overlapping aggregated series in Appendix A. By substituting (A1) from the Appendix A into (B2), we get:

$$cov(D_T, f_t) = \alpha \left( \gamma_1 \times \frac{1-\phi^h}{1-\phi} + (1-\alpha)\phi\gamma_1 \times \frac{1-\phi^h}{1-\phi} + \cdots + (1-\alpha)^{N-1}\phi^{N-1}\gamma_1 \times \frac{1-\phi^h}{1-\phi} \right) + (1-\alpha)^N \phi^{N-1}\gamma_1 \times \frac{1-\phi^h}{1-\phi}$$

$$= \frac{\alpha\gamma_1(1-\phi^h)}{(1-\phi)(1-\phi+\alpha\phi)} \times \left(1 - (\phi - \alpha\phi)^N\right) + (1-\alpha)^N \phi^{N-1}\gamma_1 \times \frac{1-\phi^h}{1-\phi} \quad \text{(B3)}$$

## Appendix C: Variance of non-overlapping aggregation forecast and the covariance of the aggregated demand and its aggregated forecast

By considering (12) and the fact that $var(D_{T-k,NOA}) = \gamma'_0$ and $cov(D_T, D_{T-k,NOA}) = \gamma'_k$ for all $k \geq 1$, the variance of the forecast for non-overlapping aggregated series is calculated as follows:

$$var(F^1_{T,NOA}) = var\left( \sum_{k=1}^{[N/m]} \beta_N (1-\beta_N)^{k-1} D_{T-k,NOA} + (1-\beta_N)^{[N/m]} F^1_{0,NOA} \right)$$

$$= \beta_N^2 Var(D_{T-1,NOA}) + (\beta_N(1-\beta_N))^2 Var(D_{T-2,NOA}) + \cdots + \left(\beta_N(1-\beta_N)^{[N/m]-1}\right)^2 var\left(D_{T-[N/m],NOA}\right) +$$

$$(1-\beta_N)^{2[N/m]} var\left(D_{T-[N/m],NOA}\right) + 2\left( \beta_N^2 (1-\beta_N)\gamma'_1 + \beta_N^2 (1-\beta_N)^2 \gamma'_2 + \cdots + \beta_N^2 (1-\beta_N)^{[N/m]}\gamma'_{[N/m]-1} \right) \quad \text{(C1)}$$

$$+ 2\beta_N (1-\beta_N)^{[N/m]} \gamma'_{[N/m]-1} + 2\left( \beta_N^2 (1-\beta_N)^3 \gamma'_1 + \beta_N^2 (1-\beta_N)^4 \gamma'_2 + \cdots + \beta_N^2 (1-\beta_N)^{[N/m]+1}\gamma'_{[N/m]-2} \right)$$

$$+ 2\beta_N (1-\beta_N)^{[N/m]+1} \gamma'_{[N/m]-2} + \cdots + 2\beta_N^2 (1-\beta_N)^{2[N/m]-3}\gamma'_1 + 2\beta_N (1-\beta_N)^{2[N/m]-1}\gamma'_0$$

By substituting (A4) in Appendix A into (C1) and simplifying it, we get:

$$var(F^1_{T,NOA}) = \frac{\beta_N \gamma'_0 \left(1-(1-\beta_N)^{2[N/m]}\right)}{(2-\beta_N)} + (1-\beta_N)^{2[N/m]} \gamma'_0 + 2\beta_N (1-\beta_N)^{2[N/m]-1}\gamma'_0$$

$$+ \sum_{i=1}^{([N/m]-1)} \frac{2\beta_N \gamma_1}{(2-\beta_N)} \left(\frac{1-\phi^m}{1-\phi}\right)^2 (1-\beta_N)^i \phi^{m(i-1)} \left(1-(1-\beta_N)^{2([N/m]-i)}\right) + \sum_{i=1}^{[N/m]-1} 2\beta_N (1-\beta_N)^{[N/m]+i-1}\gamma'_{[N/m]-i} \quad \text{(C2)}$$

The forecast of non-overlapping aggregated series at period $T$ is obtained by substituting (14) into (C2) as follows:

$$var(F^1_{T,NOA}) = \left( m\gamma_0 + \gamma_1 \left( \sum_{k=1}^{m-1} 2(m-k)\phi^{k-1} \right) \right) \left( \frac{\beta_N \left(1-(1-\beta_N)^{2[N/m]}\right)}{(2-\beta_N)} + (1-\beta_N)^{2[N/m]} + 2\beta_N (1-\beta_N)^{2[N/m]-1} \right)$$

$$+ \sum_{i=1}^{([N/m]-1)} \frac{2\beta_N \gamma_1}{(2-\beta_N)} \left(\frac{1-\phi^m}{1-\phi}\right)^2 (1-\beta_N)^i \phi^{m(i-1)} \left(1-(1-\beta_N)^{2[N/m]-i}\right) + 2\beta_N \gamma_1 \times \left(\frac{1-\phi^m}{1-\phi}\right)^2 \sum_{i=1}^{[N/m]-1} (1-\beta_N)^{[N/m]+i-1} \phi^{m([N/m]-i-1)}. \quad \text{(C3)}$$

The second element of Equation (11) is the covariance of the non-overlapping aggregated series and its forecast. It is calculated as follows:

$$cov(D_T, F^1_{T,NOA}) = cov\left(D_T, \sum_{k=1}^{[N/m]} \beta_N (1-\beta_N)^{k-1} D_{T-k,NOA} + (1-\beta_N)^{[N/m]} F_{0,NOA}\right) =$$

$$\beta_N\left(Cov(D_T, D_{T-1,NOA}) + (1-\beta_N) Cov(D_T, D_{T-2,NOA}) + \cdots + (1-\beta_N)^{[N/m]-1} Cov\left(D_T, D_{T-[N/m],NOA}\right)\right) \quad (C4)$$

$$+ (1-\beta_N)^{[N/m]} Cov\left(D_T, D_{T-[N/m],NOA}\right)$$

By substituting (A4) from the Appendix A into (C4), we get:

$$cov(D_T, F^1_{T,NOA}) = \beta_N \begin{pmatrix} \gamma_1 \times \left(\frac{1-\phi^m}{1-\phi}\right)^2 + (1-\beta_N)\phi^m \gamma_1 \times \left(\frac{1-\phi^m}{1-\phi}\right)^2 + (1-\beta_N)^2 \phi^{2m} \gamma_1 \times \left(\frac{1-\phi^m}{1-\phi}\right)^2 + \\ \cdots + (1-\beta_N)^{[N/m]-1} \phi^{m([N/m]-1)} \gamma_1 \times \left(\frac{1-\phi^m}{1-\phi}\right)^2 \end{pmatrix}$$

$$+ (1-\beta_N)^{[N/m]} Cov\left(D_T, D_{T-[N/m],NOA}\right) \qquad (C5)$$

$$= \beta_N \gamma_1 \times \left(\frac{1-\phi^m}{1-\phi}\right)^2 \left(1 + (1-\beta_N)\phi^m + \cdots + (1-\beta_N)^{[N/m]-1} \phi^{m([N/m]-1)}\right) + (1-\beta_N)^{[N/m]} \phi^{m([N/m]-1)} \gamma_1 \times \left(\frac{1-\phi^m}{1-\phi}\right)^2$$

$$= \beta_N \gamma_1 \times \left(\frac{1-\phi^m}{1-\phi}\right)^2 \left[\frac{1-(\phi^m - \beta_N \phi^m)^{[N/m]}}{(1-\phi^m + \beta_N \phi^m)}\right] + \phi^{m([N/m]-1)} (1-\beta_N)^{[N/m]} \left(\frac{1-\phi^m}{1-\phi}\right)^2 \gamma_1,$$

## Appendix D: Variance and autocovariance of overlapping aggregated series for ARMA(1,1)

$$\gamma''_0 = cov(D_{T,OA}, D_{T,OA}) = \text{var}(D_{T,OA}) = \text{var}(d_t + d_{t+1} + \cdots + d_{t+m-1}) =$$

$$m\gamma_0 + 2(\gamma_1 + \gamma_2 + \cdots + \gamma_{m-1}) + 2(\gamma_1 + \gamma_2 + \cdots + \gamma_{m-2}) + \cdots + 2\gamma_1 = m\gamma_0 + 2\gamma_1 \sum_{i=1}^{m-1}(m-i)\phi^{i-1} \qquad (D1)$$

For all $k>1$, $\gamma_{(-k)} = \gamma_{(k)}$.

$$\gamma''_1 = cov(D_{T-1,OA}, D_{T-2,OA}) = cov(d_{t-1} + d_{t-2} + \ldots + d_{t-m}, d_{t-2} + d_{t-3} + \ldots + d_{t-(m+1)}) =$$

$$(\gamma_1 + \gamma_2 + \ldots + \gamma_m) + (\gamma_0 + \gamma_1 + \gamma_2 + \ldots + \gamma_{m-1}) + (\gamma_{-1} + \gamma_0 + \gamma_1 + \gamma_2 + \ldots + \gamma_{m-2}) + \qquad (D2)$$

$$(\gamma_{-2} + \gamma_{-1} + \gamma_0 + \gamma_1 + \gamma_2 + \ldots + \gamma_{m-3}) + \ldots + (\gamma_{-(m-2)} + \gamma_{-(m-3)} + \cdots + \gamma_0 + \gamma_1) = \sum_{i=1}^{m}\sum_{j=1}^{m} \gamma_{|j-i+1|}$$

$$\gamma''_2 = Cov(D_{T-1,OA}, D_{T-3,OA}) = Cov(d_{t-1} + d_{t-2} + \ldots + d_{t-m}, d_{t-3} + d_{t-4} + \cdots + d_{t-(m+2)}) =$$

$$(\gamma_2 + \gamma_3 + \cdots + \gamma_{m+1}) + (\gamma_1 + \gamma_2 + \cdots + \gamma_m) + (\gamma_0 + \gamma_1 + \gamma_2 + \ldots + \gamma_{m-1}) + \qquad (D3)$$

$$(\gamma_{-1} + \gamma_0 + \gamma_1 + \gamma_2 + \ldots + \gamma_{m-2}) + \ldots + (\gamma_{-(m-3)} + \gamma_{-(m-4)} + \cdots + \gamma_0 + \gamma_1 + \gamma_2)$$

$$\gamma''_3 = Cov(D_{T-1,OA}, D_{T-4,OA}) = Cov(d_{t-1} + d_{t-2} + \ldots + d_{t-m}, d_{t-4} + d_{t-5} + \cdots + d_{t-(m+3)}) =$$

$$(\gamma_3 + \cdots + \gamma_{m+2}) + (\gamma_2 + \cdots + \gamma_{m+1}) + (\gamma_1 + \gamma_2 + \ldots + \gamma_m) + \qquad (D4)$$

$$(\gamma_0 + \gamma_1 + \gamma_2 + \ldots + \gamma_{m-1}) + \ldots + (\gamma_{-(m-4)} + \gamma_{-(m-5)} + \cdots + \gamma_0 + \gamma_1 + \gamma_2 + \gamma_3)$$

$$\gamma_4'' = Cov(D_{T-1,OA}, D_{T-5,OA}) = Cov(d_{t-1} + d_{t-2} + \ldots + d_{t-m}, d_{t-4} + d_{t-5} + \cdots + d_{t-(m+3)}) =$$
$$(\gamma_3 + \cdots + \gamma_{m+2}) + (\gamma_2 + \cdots + \gamma_{m+1}) + (\gamma_1 + \gamma_2 + \ldots + \gamma_m) +$$
$$(\gamma_0 + \gamma_1 + \gamma_2 + \ldots + \gamma_{m-1}) + \ldots + (\gamma_{-(m-4)} + \gamma_{-(m-5)} + \cdots + \gamma_0 + \gamma_1 + \gamma_2 + \gamma_3) \quad \text{(D5)}$$

If we continue for auto-covariance lag $k$, we get

$$\gamma_k'' = cov(D_{T-1,OA}, D_{T-(k-1),OA}) = \gamma_k + \gamma_{k+1} + \cdots + \gamma_{k+(m-1)} +$$
$$\gamma_{k-1} + \gamma_k + \cdots + \gamma_{k+(m-2)} + \cdots + \gamma_{k-(m-1)} + \gamma_{k-(m-1)+1} + \cdots + \gamma_k = \sum_{j=0}^{m-1} \sum_{i=0}^{m-1} \gamma_{|k-i+j|} \quad \text{(D6)}$$

Therefore, the autocovariance of the overlapping aggregated process can be represented as:

$$\gamma_k'' = cov(D_{T-1,OA}, D_{T-k-1,OA}) = \begin{cases} m\gamma_0 + 2\gamma_1 \sum_{i=1}^{m-1}(m-i)\phi^{i-1}, & k = 0, \\ \sum_{i=1}^{m}\sum_{j=1}^{m} \gamma_{|j-i+1|} = \sum_{\substack{i=1 \\ \forall |j-i+1|=0}}^{m}\sum_{j=1}^{m} \gamma_0 + \sum_{\substack{i=1 \\ \forall |j-i+1|=1}}^{m}\sum_{j=1}^{m} \gamma_1 + \sum_{\substack{i=1 \\ \forall |j-i+1|>1}}^{m}\sum_{j=1}^{m} \phi^{(|j-i|)}\gamma_1, & |k|=1, \\ \sum_{i=1}^{m}\sum_{j=1}^{m} \gamma_{|k+j-i|} = \sum_{\substack{i=1 \\ \forall |k+j-i|=0}}^{m}\sum_{j=1}^{m} \gamma_0 + \sum_{\substack{i=1 \\ \forall |k+j-i|=1}}^{m}\sum_{j=1}^{m} \gamma_1 + \sum_{\substack{i=1 \\ \forall |k+j-i|>1}}^{m}\sum_{j=1}^{m} \phi^{(|k+j-i|-1)}\gamma_1, & |k|>1 \end{cases} \quad \text{(D7)}$$

Equation (D7) shows the relations between the autocovariance of overlapping TA and the non-aggregated demand process.

## Appendix E: Covariance of the overlapping aggregated series and its forecast

Similar to ARMA(1,1) process, the covariance between $D_{T,OA} = d_t + d_{t+1} + \cdots + d_{t+m-1}$ and $d_{t-k}$ can be calculated as following:

$$Cov(D_{T,OA}, d_{t-k}) = \phi^{k-1}\gamma_1 \times \frac{1-\phi^m}{1-\phi} \quad , for\, k \geq 1 \quad \text{(E1)}$$

The covariance between overlapping aggregated demand at period $T$ and period $T$-$k$ is as follows:

$$cov(D_{T,OA}D_{T-k,OA}) = cov(D_{T,OA}, d_{t-k} + d_{t-k-1} + \cdots + d_{t-(k+m-1)}) =$$
$$cov(D_{T,OA}, d_{t-k}) + cov(D_{T,OA}, d_{t-(k+1)}) + \cdots + cov(D_{T,OA}, d_{t-(k+m-1)})$$
(E2)

By substituting (E1) into (E2), it reduces to:

$$Cov(D_{T,OA}, D_{T-k,OA}) =$$
$$\phi^{k-1}\gamma_1 \times \left(\frac{1-\phi^m}{1-\phi}\right) + \phi^k \gamma_1 \times \left(\frac{1-\phi^m}{1-\phi}\right) + \cdots + \phi^{k+m-2}\gamma_1 \times \left(\frac{1-\phi^m}{1-\phi}\right) =$$
$$\phi^{k-1}\gamma_1 \times \left(\frac{1-\phi^m}{1-\phi}\right)(1+\phi+\cdots+\phi^{m-1}) = \phi^{k-1}\gamma_1 \times \left(\frac{1-\phi^m}{1-\phi}\right)^2$$
(E3)

Equation (E3) gives the covariance of the overlapping aggregated series at period $T$ and aggregated demand at period $T-k$, $k>1$. The covariance of overlapping aggregated demand and its forecast is calculated as follows:

$$Cov(D_T, F^1_{T,OA}) = Cov\left(D_T, \sum_{k=1}^{N-m+1} \beta_O (1-\beta_O)^{k-1} D_{T-k,OA} + (1-\beta_O)^{N-m+1} F^1_{0,OA}\right)$$
$$= \beta_O Cov(D_T, D_{T-1,OA}) + \beta_O(1-\beta_O)Cov(D_T, D_{T-2,OA}) + \beta_O(1-\beta_O)^2 Cov(D_T, D_{T-3,OA})$$
$$+ \cdots + \beta_O(1-\beta_O)^{N-m}Cov(D_T, D_{T-(N-m+1),OA}) + Cov(D_T, (1-\beta_O)^{N-m+1} D_{T-(N-m+1)})$$
(E4)

By substituting (E3) into (E4) the covariance of overlapping aggregated demand and its forecast is:

$$Cov(D_T, F^1_{0,OA}) = \beta_O \gamma_1 \times \left(\frac{1-\phi^m}{1-\phi}\right)^2 + \beta_O(1-\beta_O)\phi\gamma_1 \times \left(\frac{1-\phi^m}{1-\phi}\right)^2 + \beta_O(1-\beta_O)^2 \phi^2 \gamma_1 \times \left(\frac{1-\phi^m}{1-\phi}\right)^2$$
$$+ \cdots + \beta_O(1-\beta_O)^{N-m}\phi^{N-m}\gamma_1 \times \left(\frac{1-\phi^m}{1-\phi}\right)^2 + (1-\beta_O)^{N-m+1}\gamma_{(N-m+1)}$$
$$= \beta_O \gamma_1 \times \left(\frac{1-\phi^m}{1-\phi}\right)^2 \left(1+\phi(1-\beta_O)+(\phi(1-\beta_O))^2 + \cdots + (\phi(1-\beta_O))^{N-m}\right) + (1-\beta_O)^{N-m+1}\phi^{N-m}\left(\frac{1-\phi^m}{1-\phi}\right)^2 \gamma_1$$
$$= \beta_O \gamma_1 \times \left(\frac{1-\phi^m}{1-\varphi}\right)^2 \left(\frac{1-(\phi(1-\beta_O))^{N-m+1}}{1-\phi+\phi\beta_O}\right) + (1-\beta_O)^{N-m+1}\phi^{N-m}\left(\frac{1-\phi^m}{1-\phi}\right)^2 \gamma_1,$$
(E5)

## Appendix F: Variance of forecast of the overlapping aggregated demand

The variance of the forecast for overlapping aggregated demand is as follows:

$$Var\left(F_{T,OA}^{1}\right) = Var\left(\sum_{k=1}^{N-m+1} \beta_{o}\left(1-\beta_{o}\right)^{k-1} D_{T-k} + \left(1-\beta_{o}\right)^{N-m+1} F_{0,OA}\right) =$$

$$Var\left(\beta_{o} D_{T-1} + \beta_{o}\left(1-\beta_{o}\right) D_{T-2} + \beta_{o}\left(1-\beta_{o}\right)^{2} D_{T-3} + \cdots + \beta_{o}\left(1-\beta_{o}\right)^{N-m} D_{T-(N-m+1)} + \left(1-\beta_{o}\right)^{N-m+1} D_{T-(N-m+1)}\right)$$

$$= \beta_{o}^{2}\left(\gamma_{0}'' + \left(1-\beta_{o}\right)^{2} \gamma_{0}'' + \cdots + \left(1-\beta_{o}\right)^{2(N-m)} \gamma_{0}''\right) + \left(1-\beta_{o}\right)^{2(N-m+1)} \gamma_{0}''$$

$$\left(2\beta_{o}^{2}\left(1-\beta_{o}\right)\gamma_{1}'' + 2\beta_{o}^{2}\left(1-\beta_{o}\right)^{2} \gamma_{2}'' + \cdots + 2\beta_{o}^{2}\left(1-\beta_{o}\right)^{N-m} \gamma_{N-m}''\right) + 2\beta_{o}\left(1-\beta_{o}\right)^{N-m+1} \gamma_{N-m}'' \quad (F1)$$

$$\left(2\beta_{o}^{2}\left(1-\beta_{o}\right)^{3} \gamma_{1}'' + 2\beta_{o}^{2}\left(1-\beta_{o}\right)^{4} \gamma_{2}'' + \cdots + 2\beta_{o}^{2}\left(1-\beta_{o}\right)^{N-m+1} \gamma_{N-m-1}''\right) + 2\beta_{o}\left(1-\beta_{o}\right)^{N-m+2} \gamma_{N-m-1}''$$

$$\cdots + \left(2\beta_{o}^{2}\left(1-\beta_{o}\right)^{2N-2m-3} \gamma_{1}'' + \left(2\beta_{o}^{2}\left(1-\beta_{o}\right)^{2N-2m-2} \gamma_{2}''\right)\right) + 2\beta_{o}\left(1-\beta_{o}\right)^{2(N-m)} \gamma_{1}''$$

$$+ \left(2\beta_{o}^{2}\left(1-\beta_{o}\right)^{2N-2m-1} \gamma_{1}''\right) + 2\beta_{o}\left(1-\beta_{o}\right)^{2(N-m)+1} \gamma_{0}''$$

By simplifying the equation (F1), we get:

$$Var\left(F_{T,OA}\right) = \frac{\beta_{o}\gamma_{0}''\left(1-\left(\left(1-\beta_{o}\right)^{2}\right)^{N-m+1}\right)}{\left(2-\beta_{o}\right)} + 2\beta_{o}\left(1-\beta_{o}\right)^{2(N-m)+1} \gamma_{0}'' + \left(1-\beta_{o}\right)^{2(N-m+1)} \gamma_{0}''$$

$$+ \frac{2\beta_{o}}{\left(2-\beta_{o}\right)} \sum_{k=1}^{N-m}\left(1-\beta_{o}\right)^{k} \gamma_{k}''\left(1-\left(\left(1-\beta_{o}\right)\right)^{2(N-m+1-k)}\right) + 2\beta_{o}\left(1-\beta_{o}\right)^{(N-m+k)} \gamma_{N-m-k+1}'' \quad (F2)$$

By substituting (D7) into (F2), the variance of overlapping aggregated series is reduce to:

$$Var\left(F_{T,OA}^{1}\right) = \left(m\gamma_{0} + 2\gamma_{1}\sum_{i=1}^{m-1}(m-i)\phi^{i}\right)\left(\frac{\beta_{o}\left(1-\left(\left(1-\beta_{o}\right)^{2}\right)^{N-m+1}\right)}{\left(2-\beta_{o}\right)} + 2\beta_{o}\left(1-\beta_{o}\right)^{2(N-m)+1} + \left(1-\beta_{o}\right)^{2(N-m+1)}\right)$$

$$+ \frac{2\beta_{o}}{\left(2-\beta_{o}\right)} \sum_{k=1}^{N-m}\left(1-\beta_{o}\right)^{k}\left(\sum_{\substack{i=1\\ \forall |k+j-i|=0}}^{m}\sum_{j=1}^{m}\gamma_{0} + \sum_{\substack{i=1\\ \forall |k+j-i|=1}}^{m}\sum_{j=1}^{m}\gamma_{1} + \sum_{\substack{i=1\\ \forall |k+j-i|>1}}^{m}\sum_{j=1}^{m}\phi^{(|k+j-i|-1)}\gamma_{1}\right)\left(1-\left(\left(1-\beta_{o}\right)\right)^{2(N-m+1-k)}\right) \quad (F3)$$

$$+ 2\beta_{o}\sum_{k=1}^{N-m}\left(1-\beta_{o}\right)^{(N-m+k)}\sum_{\substack{i=1\\ \forall |N-m-k+j-i|=0}}^{m}\sum_{j=1}^{m}\gamma_{0} + \sum_{\substack{i=1\\ \forall |N-m-k+j-i|=1}}^{m}\sum_{j=1}^{m}\gamma_{1} + \sum_{\substack{i=1\\ \forall |N-m-k+j-i|>1}}^{m}\sum_{j=1}^{m}\phi^{(|N-m-k+j-i|-1)}\gamma_{1},$$

# Appendix G: Ratio of MSEs for smoothing constant values of 0.1, 0.2 and 0.3

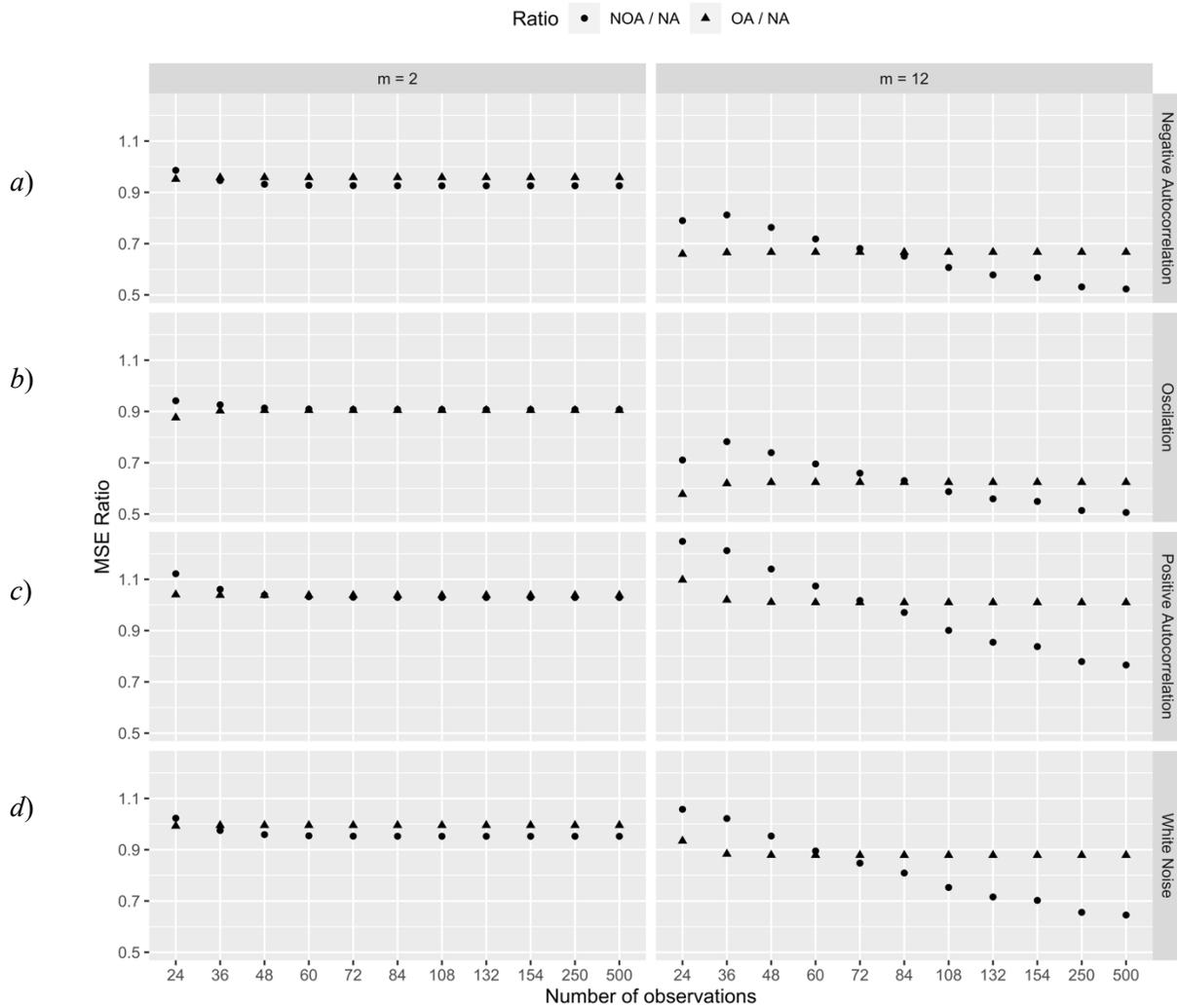

Figure G1: Ratio of MSEs for non-overlapping and overlapping aggregation to the non-aggregation approach for an ARMA(1,1) process and smoothing constant = 0.1 for all approaches with (a) negative autocorrelation, (b) autocorrelation oscillating between positive and negative, (c) positive autocorrelation and (d) no autocorrelation/white noise.

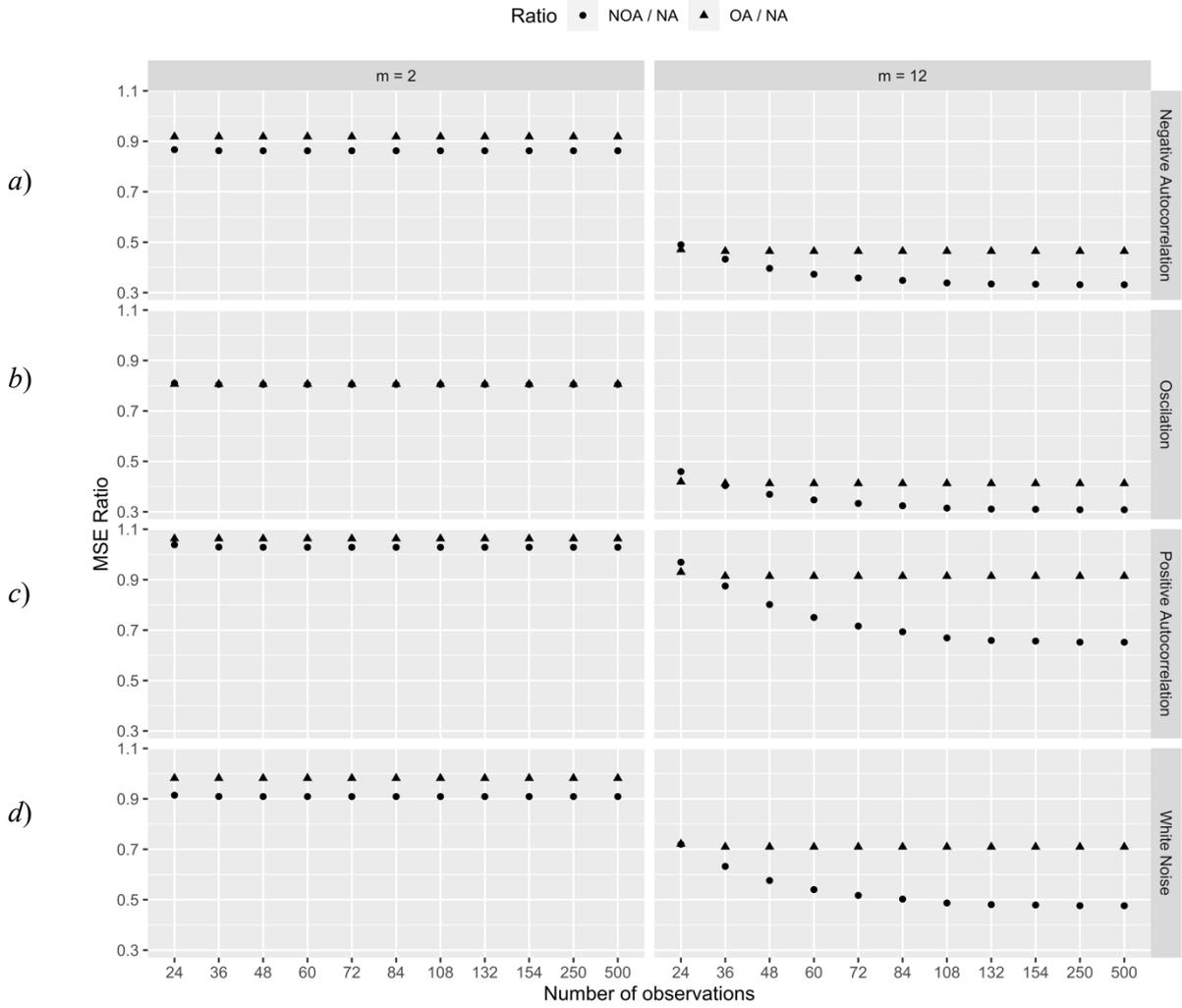

Figure G2: Ratio of MSEs for non-overlapping and overlapping aggregation to the non-aggregation approach for an ARMA(1,1) process and smoothing constant = 0.2 for all approaches with (a) negative autocorrelation, (b) autocorrelation oscillating between positive and negative, (c) positive autocorrelation and (d) no autocorrelation/white noise.

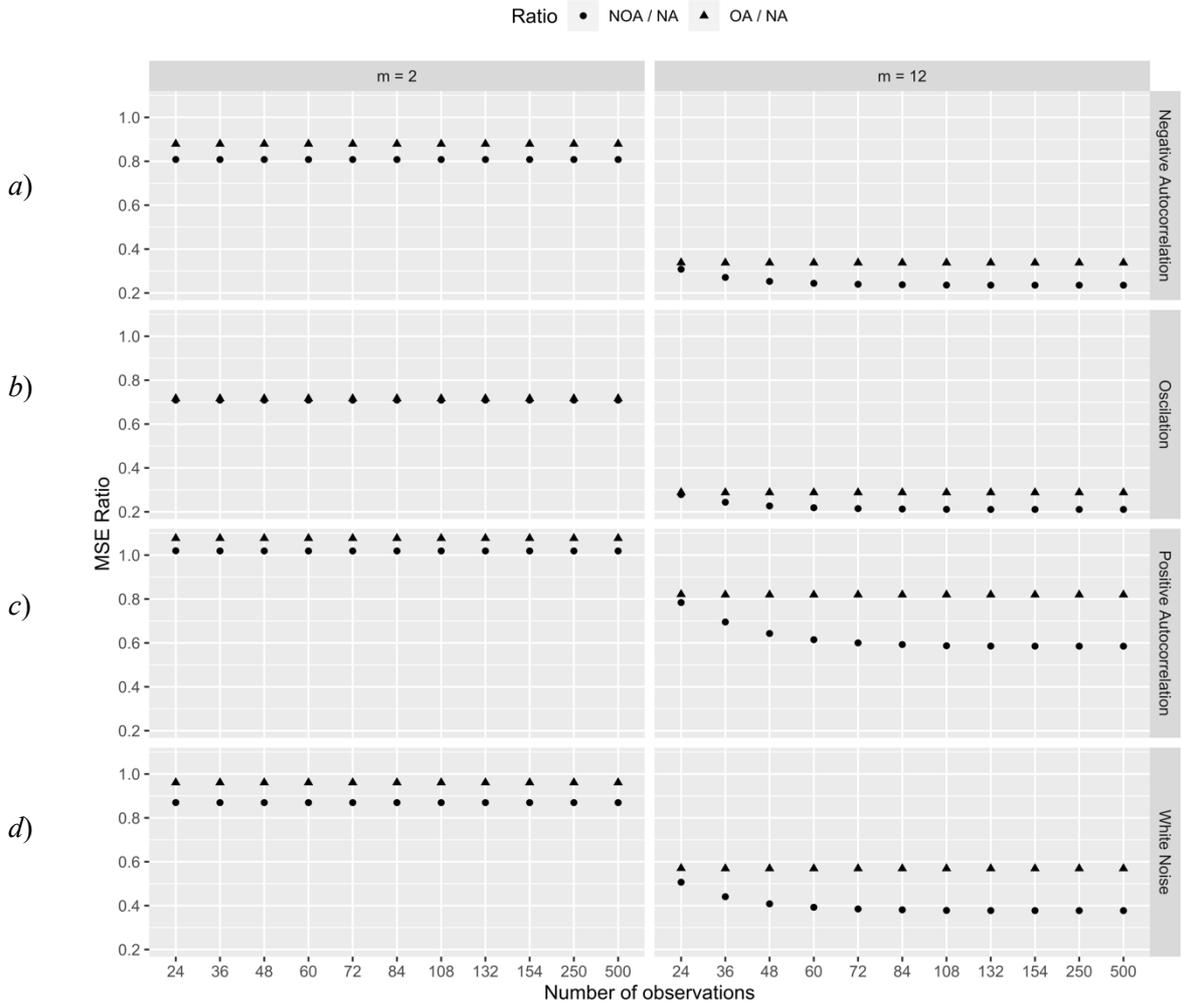

Figure G3: Ratio of MSEs for non-overlapping and overlapping aggregation to the non-aggregation approach for an ARMA(1,1) process and smoothing constant = 0.3 for all approaches with (a) negative autocorrelation, (b) autocorrelation oscillating between positive and negative, (c) positive autocorrelation and (d) no autocorrelation/white noise.